\documentclass[journal]{IEEEtran}
\usepackage{cite}
\usepackage{amsmath,amssymb,amsfonts}
\usepackage{amsthm}
\usepackage{nccmath}
\usepackage{fancyhdr}
\usepackage{setspace}
\usepackage{svg}
\usepackage{bbm}
\usepackage{soul}
\usepackage{subfloat}
\usepackage{graphicx}
\usepackage{textcomp}
\usepackage{xcolor}
\usepackage{kotex}
\usepackage{algorithmicx,algpseudocode}
\usepackage{subfigure}
\usepackage{booktabs}
\usepackage{soul,color}
\usepackage{bbold}
\usepackage{mwe}
\usepackage{multicol}
\usepackage{multirow}
\usepackage[normalem]{ulem}
\usepackage[linesnumbered,ruled]{algorithm2e}

\newcommand{\BfPara}[1]{{\noindent\bf#1.}\xspace}

\newcommand{\revision}[1]{{\color{black}{#1}}}

\begin{document}

\title{Quantum Multi-Agent Reinforcement Learning for Autonomous Mobility Cooperation}

\author{
    Soohyun Park, 
    Jae Pyoung Kim, 
    Chanyoung Park, 
    Soyi Jung, and
    Joongheon Kim,~\IEEEmembership{Senior Member, IEEE}  
    \thanks{Soohyun Park, Jae Pyoung Kim, Chanyoung Park, and Joongheon Kim are with the Department of Electrical and Computer Engineering, Korea University, Seoul 02841, Republic of Korea (e-mails: \{soohyun828,paulkim436,cosdeneb,joongheon\}@korea.ac.kr).}
    \thanks{Soyi Jung is with the Department of Electrical and Computer Engineering, Ajou University, Suwon 16499, Republic of Korea (e-mail: sjung@ajou.ac.kr).}
}
\maketitle
\begin{abstract}
For Industry 4.0 Revolution, cooperative autonomous mobility systems are widely used based on multi-agent reinforcement learning (MARL). However, the MARL-based algorithms suffer from huge parameter utilization and convergence difficulties with many agents. To tackle these problems, a quantum MARL (QMARL) algorithm based on the concept of actor-critic network is proposed, which is beneficial in terms of scalability, to deal with the limitations in the noisy intermediate-scale quantum (NISQ) era. Additionally, our QMARL is also beneficial in terms of efficient parameter utilization and fast convergence due to quantum supremacy. Note that the reward in our QMARL is defined as task precision over computation time in multiple agents, thus, multi-agent cooperation can be realized. For further improvement, an additional technique for scalability is proposed, which is called projection value measure (PVM). Based on PVM, our proposed QMARL can achieve the highest reward, by reducing the action dimension into a logarithmic-scale. Finally, we can conclude that our proposed QMARL with PVM outperforms the other algorithms in terms of efficient parameter utilization, fast convergence, and scalability.
\end{abstract}

\IEEEpeerreviewmaketitle

\section{Introduction}
One of main themes in Industry 4.0 Revolution is the pervasive autonomy of distributed intelligent computing systems, e.g., autonomous mobile robots (AMRs), unmanned aerial vehicles (UAVs), and drones, in large-scale wide-area management for smart city applications, i.e., \textit{smart factory management} and \textit{aerial cellular access}. 
For \textit{smart factory management}, the Verizon Report states, ``\textit{Industry 4.0 is squarely underway in manufacturing. The global market is expected to reach \$219.8 billion by 2026, and autonomous mobile robots are becoming key workhorses in this transformation}"~\cite{verizon}. 
Furthermore, for \textit{aerial cellular access}, Analysys Mason's Report says, ``\textit{Cellular connected drones are set to become a major growth segment in the drones market. The total revenue generated by cellular connected drones worldwide will grow to \$8\,billion in 2030}"~\cite{mason}.  

In these applications, the management of this type of large-scale wide-areas is extremely cost-ineffective if no automation is carried out by unmanned mobile computing platforms. Furthermore, multiple autonomous mobile platforms should be utilized for taking care of large-scale wide-areas that cannot be covered by a single autonomous mobile platform. 
Therefore, in order to realize the efficient multiple autonomous control mobility systems, multi-agent algorithms such as multi-agent reinforcement learning (MARL) are essentially required~\cite{iotj2010kwon}.
Based on this, various MARL-based algorithms have been designed and implemented for autonomous mobility systems. 

However, the use of conventional MARL-based algorithms is not always possible due to \revision{large parameter utilization, iterative large-delay training, convergence issues with large number of agents}. In order to tackle these problems, a new emerging approach for designing MARL-based algorithms, i.e., quantum multi-agent reinforcement learning (QMARL), is proposed in this article. By utilizing quantum computing (QC) capabilities on top of MARL, it is beneficial in terms of \revision{efficient parameter utilization and fast convergence}. 
However, due to the the limited number of qubit utilization in noisy intermediate-scale quantum (NISQ) era, the scale of the neural network (NN) for training quantum reinforcement learning models cannot be easily extended. Therefore, a new design especially for quantum reinforcement learning is needed. In this article, a scalable quantum actor-critic networks for training multiple quantum-capable agents under the limitations of the number of qubits in the NISQ era will be introduced.  
After designing a novel quantum actor-critic networks, the applications of the networks are discussed in terms of (i) smart factory management via AMR, and (ii) mobile cellular access via UAVs. The overview of each given scenario will be explained and the performance evaluation will be conducted. Based on the evaluation results, the practicality of quantum supremacy for autonomous mobility systems and algorithms will be corroborate.


\section{Autonomous Mobility Systems and Algorithms}\label{sec:2}
\subsection{Autonomous Mobility Cooperation Systems}
In modern communications and networking research, the use of autonomous mobility systems is widely discussed~\cite{tii202210yun, network22ren}. Among various autonomous mobility systems, UAV and AMR are actively utilized in various learning-based distributed computing applications~\cite{tii202210yun, iotj2023yun}. 
The reasons why the UAV and AMR systems are practically utilized in many applications can be categorized as follows. 
\begin{itemize}
    \item \BfPara{Accessibility} Due to the advancement of manufacturing technology, the cost of producing such UAVs and AMRs has decreased significantly in recent years. Consequently, the access to these technologies has become widely accessible and cheaper which is why the practical utilization of UAV and AMR systems is beneficial.
    \item \BfPara{Flexibility} Both UAVs and AMRs are small and capable of extensive mobilization to remote places. Therefore, these technologies can provide its services to people who are far away from urban areas. Furthermore, as these agents can reach locations that are inaccessible to humans, e.g., underwater, radioactive, they can prevent loss of human lives by carrying out dangerous jobs.
    \item \BfPara{Usability} The emergence of state-of-the-art technologies, e.g., cameras, robotic arms, high-power motors, have allowed UAVs and AMRs to execute increasingly complex tasks. Thus, the range of tasks that can be automated by these technologies is increasing even now which proves the high usability of the two technologies. 
\end{itemize}

\subsection{Autonomous Mobility Cooperation Algorithms}
In order to control autonomous mobility systems, various algorithms have been widely proposed. 
For realizing the autonomous control of multiple agents, it is essential to design and implement algorithms based on artificial intelligence (AI) and \revision{machine} learning methods. 
Among various AI and \revision{machine} learning methods, reinforcement learning (RL) is mainly considered for autonomous mobility control algorithm design and implementation because of the formal definition of the RL-based algorithms as discrete-time stochastic decision control under observations~\cite{pieee202105park}.
In addition, the RL-based algorithms have better adaptability to environmental uncertainties, e.g., pedestrians, obstacles, or traffic situations, in comparison to other AI and \revision{machine} learning algorithms as they sequentially perform run-time decision-making. Thus, RL is indeed a suitable algorithm for autonomous mobility control because autonomous mobility agents have to consistently deal with unexpected situations. 
In RL, the agents take actions based on the given state in order to maximize the expectation of accumulative rewards. 
%
\revision{Especially for UAVs in mobile access applications, RL-based algorithms offer valuable benefits, empowering these aerial platforms to autonomously optimize their actions based on state information such as battery levels and altitudes for 3D trajectory. This facilitates informed decision-making and efficient resource utilization, which is crucial for achieving effective and sustainable UAV operations.
Furthermore, for AMRs in smart factory applications, the utilization of RL-based algorithms can enhance their capabilities to navigate intricate factory layouts, avoid obstacles effectively, and optimize their given task performance within complex industrial settings.}
%
\revision{Here, RL-based UAVs with limited battery capacity can consider their own current energy levels and altitudes as state information. By assigning negative rewards, i.e., penalties, when the batteries are depleted or when the altitudes exceed certain thresholds based on their observed information, UAVs can learn policies that guide their actions to maintain appropriate energy states as well as avoid excessively high altitudes.
Similarly, RL-based AMRs can encounter obstacles during indoor path planning. In this case, the agents can perceive the obstacles observed in the environment as state information, allowing them to learn parameters during policy training.
By assigning penalties when agents collide with obstacles during the training phase, the agents enable to adjust the parameters of their policies in a manner that encourages obstacle avoidance whenever the obstacle is detected.
In summary, it can be concluded that RL-based algorithms exhibit significant advantages in sequential decision-making scenarios such as the network control and management bu multiple mobile agents, i.e., multi-agent autonomous mobility cooperation, while adapting to dynamic environmental uncertainties.}
\revision{For more details}, in order to control multiple mobile agents, a single-agent RL architecture is not appropriate. 
Suppose multiple RL agents with identical sets of action-reward pairs exist. Since all the agents share the same action and reward set, they will collectively react in an identical manner to a given situation. This phenomenon is not desirable in a multi-agent RL model because an efficient system requires the agents to be organic and produce a variety of distinct action sets focused on solving a formulated singular problem of the environment. 
Therefore, when multiple RL agents are deployed, they are expected to be coordinated and cooperated with each other.
For this purpose, information about the environment collected by each agent must be distributed to all other agents to implement MARL~\cite{tii202210yun}. 
Until now, many MARL algorithms such as the Communication Network (CommNet) model has been proposed and actively used in many distributed computing applications~\cite{sukhbaatar2016learning}.
In CommNet, a single NN is defined where the inputs and outputs are the combinations of multiple agents' states and actions. As a result, information sharing can be achieved when the combined multi-agent information is processed in a single NN. After training of the single NN is finished, it will be distributed to multiple agents for simultaneous and distributed execution. Then, each mobility device such as UAV and AMR acting as an RL agent can conduct its own inference with the trained NN which has the information of all given RL agents.
\revision{In addition to design the NN architecture based on the number of agents for facilitating the collaboration in multi-agent systems such as CommNet, the other algorithm, i.e., a \textit{centralized critic and multiple actor network}, can be also one of promising approaches~\cite{iotj2023yun}. This approach leverages a centralized critic that evaluates the policies of multiple actors (i.e., agents) in order to realize cooperative decision-making among them without increasing input dimension due to the number of agents.}
\revision{These} concept\revision{s for MARL, i.e., CommNet and actor-critic,} \revision{are generally} called centralized training and distributed execution (CTDE).

\begin{figure*}[!t]
    \centering
    \includegraphics[width=\linewidth]{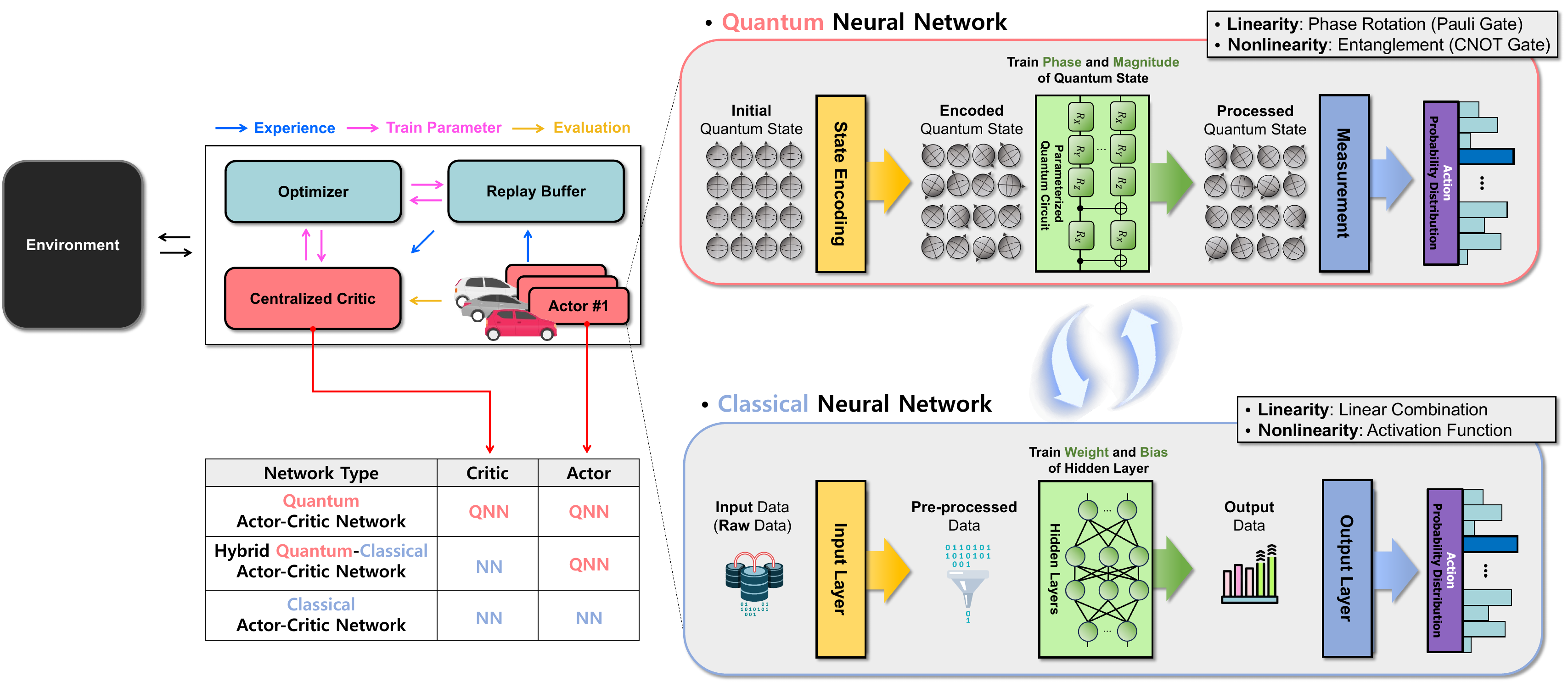}
    \caption{Structure of our proposed quantum actor-critic network and generalized quantum neural networks.}
    \label{fig:gqnn}
\end{figure*}

\section{Quantum Reinforcement Learning}\label{sec:3}

\subsection{Quantum Supremacy and Limitations}\label{sec:3a}
\subsubsection{Quantum Supremacy}
The concept of quantum supremacy was clearly demonstrated by \textit{Google} in 2019~\cite{arute2019quantum}. Via this work, the feasibility of full-scale QC that outperforms the classical supercomputer in terms of computational cost and time has been proven. Considering how classical computing power is reaching its upper bound, this progress holds significant meaning in machine learning research because the unique strengths of quantum computers have the potential to open new horizons for various research areas, e.g., reinforcement learning, pattern recognition, and pre-trained natural language processing. By exploiting the advantages of quantum supremacy, problems that could not be solved by classical supercomputers can now be overcome~\cite{shor1995scheme}. 
Essentially, the strength of QC originates from the difference in the basic unit of information used. In QC, \textit{qubits} are used instead of the familiar bits of classical computing. While both operate between the values of 0 and 1, qubits are not deterministic unlike bits. Bits can only be either 0 or 1 at any point of time but qubits can exist as two values simultaneously because they are expressed by the probability amplitudes of the respective values. This unique trait of qubits is known as the state of \textit{superposition} which allows $n$ qubits to express $2^n$ amount of information. Therefore, as the information density in each qubit is higher than each bit, quantum computing can express the same amount of information while using lesser parameters than classical computing. With this characteristic, QC can not only process more data but also perform computations more quickly. Furthermore, QC can also be used to build quantum neural networks (QNNs) which can replace classical NNs by using qubits and quantum logic gates~\cite{lockwood20}.
Although both QNNs and NNs have the same function, QNNs are relatively easier to implement and requires lesser computing resources than NNs because of the smaller number of parameters used.
\revision{In summary, the novelties and advantages of QNNs comparing to NNs can be categorized as follows.
\begin{itemize}
    \item \revision{\BfPara{Parameter Utilization} QNNs have the ability to achieve similar performance to conventional NNs with fewer parameters, primarily due to superposition and entanglement. The superposition allows the quantum system to exist in multiple states at once, and thus, it expands the representational capacity of QNNs. In addition, the entanglement provides robust interconnection among quantum states, and thus, it can realize complex and nonlinear feature representation with fewer parameters.}

    \item \revision{\BfPara{Convergence} In QNNs, the \textit{parameter shift rule} is used for parameter optimization during training~\cite{iotj23park}. Relative to the backpropagation used in traditional NNs, the \textit{parameter shift rule} offers simplistic and direct approaches. Therefore, the training can be accelerated in QNNs.} 

\end{itemize}
}

\subsubsection{Limitations to QC/QMARL in the NISQ Era}
\revision{On the other hand, severe limitations exist for QC and even for QMARL as well. Among them, the NISQ characteristic is the most crucial limitation and it refers to the lack of quantum error correcting technologies. Although QC/QMARL can process a large amount of data, it also has to correct an equally numerous number of errors occurring during computation. As a result of these two phenomena, the number of input qubits that can be used in a QMARL model is extremely restricted, which leads to poor performance. These error and scale related issues cannot be resolved in modern QC and QMARL, and currently developing QMARL algorithms are designed and implemented in order to achieve desired performance while experiencing the limitations.}

\subsection{Quantum Actor-Critic Networks}\label{sec:III-B}
In this section, the overall process of the quantum actor-critic network and the structure of a generalized QNN as shown in Fig.~\ref{fig:gqnn} will be elaborated. All actors and critic in this system are implemented with QNN. Therefore, to understand how training is carried out, the mechanism of QNN must be familiarized. 

\subsubsection{Quantum Neural Network} 
The most generally used QNN today is comprised of three main parts which are \revision{state} encoding, \revision{parameterized quantum} circuit \revision{(PQC)}, and measurement, \revision{as illustrated in Fig.~\ref{fig:gqnn}}.
Firstly, the state encoding converts classical input data into \revision{encoded} quantum states because bit data cannot be used with quantum circuits. \revision{This corresponds to the input layer in conventional NNs.}
Since most of the data collected today are expressed in classical bits, this encoding process is necessary to ensure compatibility between QNN and classical data. Moreover, the encoding unit is made up of several basic quantum rotation gates which manipulate single qubits.
Then, the quantum state data produced by the encoding unit reaches \revision{PQC} which is the essential part responsible for imitating the effect of a classical NN. In this layer, computation is performed on the quantum data by a set of \revision{Pauli gates ($R_X$, $R_Y$, $R_Z$) that conduct phase rotations on the quantum state along the $x$-, $y$-, and $z$-axes in 3D Bloch sphere to represent linearity} and \revision{controlled-NOT (CNOT)} gates such that the desired output can be produced.
\revision{Here,} \revision{CNOT} gates \revision{make nonlinearity} \revision{by causing} quantum entanglement between two or more qubits which is equivalent to the \textit{fully connected} structure \revision{and the activation function} of the classical NN. Thus, these gates are important in emulating the effects of NN and the parameters of these quantum gates are the values that must be trained and optimized over many iterations, resulting in better model performance.
Lastly, the output quantum data reaches the measurement unit which is paramount in QC. \revision{It serves as the counterpart to the output layer in the traditional NN, responsible for generating the final outputs.} Before measurement, quantum states are unstable because of qubits in superposition states. As a result, quantum data cannot be used for computation unless the qubits are stabilized. Hence, this is achieved by measuring the quantum states which is equivalent to performing projection on the quantum states. Furthermore, a quantum state that has been measured is known as an \textit{observable}.

Based on this introduction to QNN design, the process of training QNN will be explored and the difference between the training of classical NN will be also highlighted~\cite{network21kim}. As mentioned above, computations cannot be performed on qubits which means that backpropagation via differentiation is impossible. This is mainly due to differentiation operation causing an unwanted shift in quantum states, producing an output that is different from the desired output. 
Yet, in order to train QNN, backpropagation must be implemented in a way such that the quantum gradient can be calculated while still maintaining the integrity of the quantum states. Hence, the \textit{parameter shift rule}~\cite{mitarai2018quantum} is used which expresses the derivative of a quantum state in linear combination. Via this method, qubits can be differentiated and QNNs can be trained by computing quantum gradients.

\begin{figure*}[!t]
    \centering
    \includegraphics[width=\linewidth]{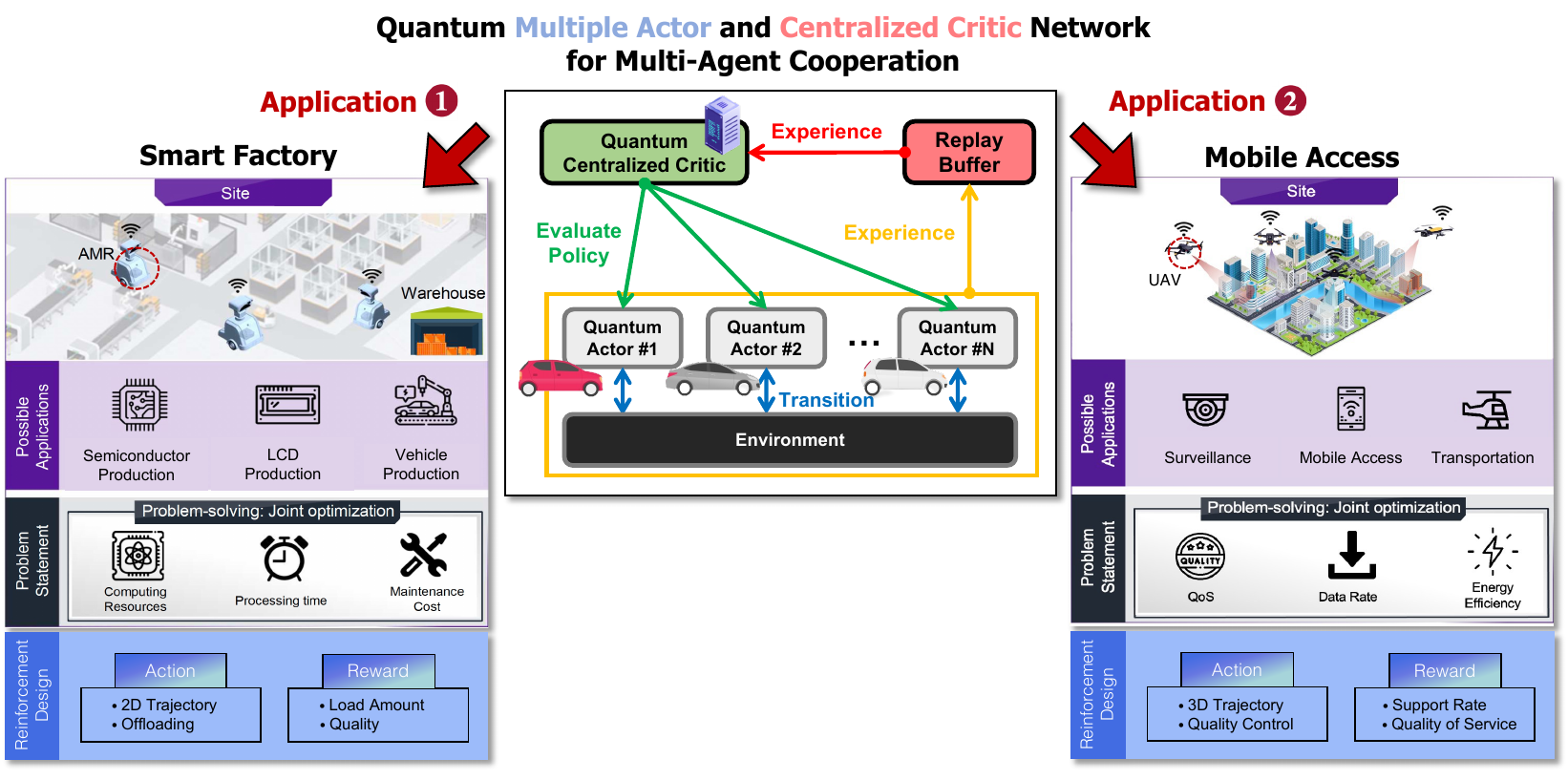}
    \caption{\revision{Various autonomous multi-agent mobility cooperation applications using quantum multiple actors and centralized critic network.}}
    \label{fig:applic}
\end{figure*}

\subsubsection{Quantum Actor-Critic Network}
Next, the overall process of the quantum actor-critic network system is investigated.
%
\revision{As depicted in Fig.~\ref{fig:gqnn}, based on how the networks for the actor and critic are constructed, either using traditional NN or QNN, the actor-critic network can be categorized into three types: \textit{i)} quantum actor-critic network, \textit{ii)} hybrid quantum-classical actor-critic network, and \textit{iii)} classical actor-critic network.}
In the\revision{se} \revision{networks}, \revision{multiple} actors execute specific actions according to their given \revision{states in the environment} by utilizing their policies. As a result of these actions, corresponding states, rewards, and observations are created by each agent which are then transmitted to the replay buffer \revision{that} acts as a storage unit of these sets of information and supplies batches of data to the \revision{centralized} critic.
Next, in the centralized critic, the data about the experiences of actors are received and used to train its NN parameters as well as produce the estimated value function which is used as a training guide for the actors. Therefore, the local actors will be trained with the given value function. After the update is completed in each actor, the updated policies will be used to infer a set of action distribution in which all the actions of the actors are included. By executing these actions, another iteration of the aforementioned process will begin. 
As a result, both actors and critic will be trained \revision{with optimizer} until the optimal policies of the actors are finally obtained.

\revision{
\subsubsection{Benefit of the Actor-Critic Network Design in NISQ-Limited QC/QMARL}
As described in Sec.~\ref{sec:2}, CommNet~\cite{sukhbaatar2016learning} is an architecture that takes the state information of all agents as an input, and trains their corresponding actions using a single NN. In this NN of CommNet, the input size escalates to $N\cdot M$ when $N$ and $M$ stand for the number of agents and the dimension of state spaces, respectively. Consequently, the input size experiences a linear scale-up commensurate with the increment in the number of agents. However, such linear scaling of input dimension according to the increase of the number of agents poses challenges for reward convergence in conventional CommNet-based MARL settings, particularly when dealing with environments populated with a massive number of agents. Such a phenomenon is equally applicable in QNNs, presenting substantial burdens and scalability issues in QMARL during the NISQ era. 

In this study, to circumvent such as the challenges due to the limited number of utilized qubits, a centralized critic and multiple actor network, as illustrated in Fig.~\ref{fig:gqnn}, is employed to cooperate with multiple agents and ensure that the input dimension size of multiple agents remains independent to the number of agents. In summary, this approach can be obviously beneficial because the input sizes of QNNs of critic and actors are not scaled up with the number of agents. Therefore, this approach superiors in QC-based multi-agent algorithm design, comparing to CommNet, during the modern NISQ era.
}

\section{Autonomous Mobility Applications}\label{sec:4}

\subsection{Various QMARL Applications for Autonomous Mobility}

\BfPara{Scenario Overview of Smart Factory \revision{Application}}
This section discusses the application of QMARL in smart factory environment\revision{~\cite{iotj2023yun}}.
\revision{Generally, AMRs are commonly employed for tasks such as transporting goods within the factory premises or actively participating in manufacturing process (e.g., semiconductor or vehicle production) to enhance task performance while reducing processing time and maintenance costs.}
Therefore, numerous essential tasks must be assigned to the AMRs to ensure that the smart factory is properly operated.
For instance, in \revision{autonomous liquid crystal display} \revision{(}LCD\revision{)} \revision{manufacturing} factory, these AMRs must transport LCDs to designated warehouses for storage.
During transportation, the LCDs must be handled very carefully as any form of impact on the fragile LCDs will cause damage.
Hence, the maximum capacities of the AMRs cannot be exceeded as attempting to transport excessive number of LCDs will cause the AMRs to drop the LCDs.
Furthermore, the overloading of the AMRs will increase the probability of the AMRs experiencing mechanical malfunction which should be avoided because it will incur repair costs.
Next, the warehouse capacity refers to the remaining storage space available and the overloading of warehouse will hinder the operation of the smart factory because there will not be enough space to store the produced LCDs.
Therefore, queues of the two components regarding load amounts must be managed via QMARL such that the smart factory can be operated normally.
At the same time, the AMRs are also responsible for the quality control of the manufactured LCDs.
Among all the LCDs produced, there may exist some faulty products unfit for the usage which must be differentiated from the properly produced batch of LCDs.
Thus, the AMRs utilize a parameter known as precision to decide if an LCD is suitable for sale to customers.
In summary, AMRs can perform optimal path planning and offloading by setting a reward function that maintains high quality and appropriate load amounts.
The main problem of this scenario can be formulated as a queue management problem where the queues of AMRs and warehouse represent the maximum capacities of each component.
Furthermore, this queue management problem can be applied to a wide range of other smart factory applications because if the tasks of AMRs in the LCDs are changed appropriately, the LCD factory scenario can be easily applied to other environments.
Thus, by analyzing the queue management problem, other applications can be investigated as well.

\BfPara{Scenario Overview of Mobile Access Application}
This section discusses the application of QMARL in the aerial cellular access environment. 
In~\cite{iotj23park}, an autonomous mobile access network with multiple UAVs is proposed aiming to provide reliable wireless communication services to ground users, with a focus on delivering high quality of service (QoS) and fast data rates.
The mobile access system can be applied to efficient surveillance or transportation services without the constraint of physical location.
UAVs can leverage their state information, including battery status, altitude, and obstacle detection, to plan energy-efficient and optimal 3D paths with quality control. Their decision-making capabilities can be guided by a reward function that aims at maximizing support rate for performance improvement.
Similar to other models, a CTDE\revision{-based} quantum actor-critic network is utilized in this application as well. To train their policies, the UAVs will send their experience data to the cloud server. Furthermore, two types of noises, i.e., state and action noises, are additionally considered to reflect more realistic environments. The former component refers to the noise of global positioning system (GPS) receivers, i.e., interference, jamming, and delay lock loop, which occurs when determining the locations of UAVs using GPS. During the process of sending positions of UAVs to the central cloud server, there may be noise errors for GPS sensors. The action noise corresponds to the influence of wind on the data of UAV. Since wind affects the position of UAVs, it interferes with the coherence of UAV data which can affect the performance of the model. 
In summary, the CTDE-based QMARL training helps multi-UAV to cooperate and overcomes the scalability issue regarding qubits. Thus, the possibility of mounting actual large-scale quantum computers on UAVs does not have to be considered as only two to ten qubits are required. 

\begin{figure}[!ht]
    \centering
    \includegraphics[width=\linewidth]{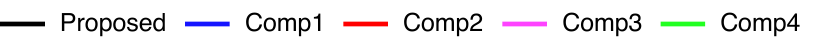}\\
    \includegraphics[width=.9\linewidth]{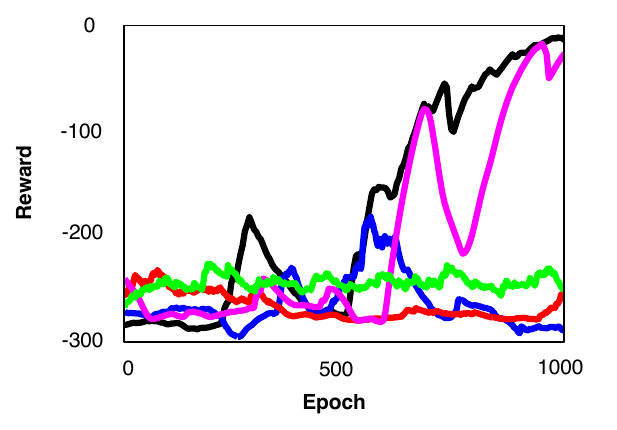}
    \caption{\revision{Simulation Results -- Total Reward.}}
    \label{fig:results}
\end{figure}

\begin{table}[!ht]
\caption{\revision{Comparative Analysis of Parameter Settings in Benchmarks}}
\centering
\begin{tabular}{l||r|r}\toprule[1pt]
    \textbf{Schemes} & \textbf{Computing method} &\textbf{\# of parameters} \\
    \midrule
    \textbf{Proposed} & Quantum & $\approx110$\\
    Comp1 &  Quantum\,/\,Classical & $\approx110$\\
    Comp2 &  Classical & $\approx110$\\
    Comp3 &  Classical & $\approx40$K\\
    Comp4 &  Random Walk & None\\
    \bottomrule[1pt]
\end{tabular}
\label{tab:3}
\end{table}

\subsection{Performance Evaluation}
The comparison between the different scenarios has been highlighted in Fig.~\ref{fig:applic}.
\revision{The reward function for generalized performance evaluation was designed such that agents can obtain higher rewards when they demonstrate high task precision within shorter task processing time~\cite{iotj2023yun}. The evaluation results in a simulated environment with \textit{torchquantum} libraries are in Fig~\ref{fig:results}.}
\revision{An empirical evaluation juxtaposes the performance of the proposed QMARL-based framework with four other benchmark frameworks that adopt conventional MARL algorithms or random walk simulation. These benchmark frameworks are designed to verify the distinctive advantage of QMARL in operating with fewer parameters compared to existing MARL, factoring in the number of parameters involved as presented in Table~\ref{tab:3}.}
\textit{Comp1} represents the hybrid quantum-classical MARL framework utilizing 110 parameters which is the same number as the proposed framework.
\textit{Comp2} refers to the classical MARL framework utilizing 110 parameters where the number is equivalent to the one of our proposed framework and \textit{Comp1}.
Similar to \textit{Comp2}, \textit{Comp3} is also a classical MARL framework whereas it has much larger parameters, i.e., 40,000, which definitely affects the performance of this framework.
Finally, \textit{Comp4} is a classical random walk simulation which does not utilize any parameters.
Fig.~\ref{fig:results} shows the dynamics in total rewards over 1,000 epochs. Based on this result, it can be confirmed that the proposed framework achieves the highest reward and the fastest convergence compared to other benchmark frameworks. 
The conventional MARL framework ac desired performances only when large number of parameters can be utilized as shown in Table~\ref{tab:3}.
It means that the result demonstrates that the multiple agents controlled by the proposed QMARL-based quantum actor-critic network framework was able to carry out its given task with the highest precision rate based on the definition of the reward function. 
\revision{Furthermore, it is also demonstrated that the proposed QMARL-based framework is able to reduce the initial task processing time to a minimum comparing to the others.}
Thus, via this performance evaluation result, it can be verified that the proposed QMARL framework \revision{with fewer parameters} has outperformed the other benchmark frameworks, corroborating quantum supremacy.

\begin{table}[!t]
    \centering
    \caption{\revision{Average Rewards with Various Action Dimensions}}
    \begin{tabular}{l||c|c|c|c}
    \toprule
    Action Dim & \textbf{QMARL} & {MARL} & {IQL} & {Random}\\
    \midrule
    $ |\mathcal{A}| = 2^{1}$ & 0.961 & $\mathbf{1}$ & 0.928 & 0.526 \\
    $ |\mathcal{A}| = 2^{4}$ & 0.778 & $\mathbf{1}$ & 0.472 & 0.106 \\
    $ |\mathcal{A}| = 2^{16}$ & $\mathbf{1}$ & 0.137 & 0.203 & 0.301 \\
    \bottomrule
    \end{tabular}
    \label{tab:average_reward}
\end{table}

\section{\revision{Open Discussions}}
\subsection{\revision{Logarithmic-Scale Dimension Reduction for Scalability}}
\revision{
This section delves into the quantum supremacy in terms of \textit{scalability}.
In real-world environments, many problems encountered, such as combinatorial optimization or scheduling problems, often involve agents having to deal with a significantly large dimension of actions.
However, the efficiency of MARL-based approach can be significantly compromised by the high-dimensionality dilemma, known as the \textit{curse of dimensionality}. Therefore, the convergence cannot be achieved when the number of agents is massive.
%
%
To address this problem, projection value measure (PVM)-based QMARL presents a formidable solution~\cite{yun2022projection}.
This approach has the potential to streamline the decision-making process of the agents by reducing its action dimensions on a logarithmic-scale, effectively mitigating computational burdens and enhancing efficiency by avoiding quantum error occurring due to large number of utilized qubits.
%
This paper compares the performance of the proposed QMARL algorithm with the benchmarks including conventional MARL, independent Q-learning (IQL), and random algorithms.
%
%
As presented in Table~\ref{tab:average_reward}, the data underscores that the conventional MARL-based approach outperforms its counterparts across all benchmarks when the number of actions is relatively small ($2^1$ and $2^4$).
%
%
When the action space is with a higher dimension ($2^{16}$) which can well-suited for real-world scenarios, the proposed QMARL-based approach utilizing PVM prevails in comparison.
This is due to the utilization of PVM, which allows for reducing the action dimension of $2^{16}$ to a logarithmic-scale of $16$.
It means that multiple agents are capable to solve complex combinatorics optimization problems using only $16$ qubits.
As a result, it can be noted that our QMARL with PVM can be especially beneficial when large-scale actions should be considered in realistic problems which cannot be handled by conventional MARL approaches.  %
In summary, we can confirm that utilizing our QMARL-based approach with PVM is one of promising solutions to deal with the constraints in the NISQ era.}

\subsection{\revision{Potential QMARL Applications}}

\revision{According to quantum supremacy, our proposed quantum actor-critic network algorithm can be widely and actively used for many applications in addition to AMRs and UAVs cooperation. Firstly, the expansion of action dimensionality enhances QMARL's capabilities, enabling effective training of multiple agents in complex action-oriented games. This empowers agents to acquire adaptive strategies capable of dynamically responding to changes in game dynamics, thereby providing a more realistic and challenging game-play experience. Additionally, massive scale of RL can be realized only with PVM-based QMARL, i.e., large-scale of caches/edges or surveillance can be one of the potential applications. 

Furthermore, our proposed quantum actor-critic network algorithm demonstrates the ability to compute with a relatively small number of parameters, even when confronted with memory constraints imposed by the limited capacity of cube-satellites and battery/memory-limited Internet-of-Things (IoT) devices due to space and weight limitations. Consequently, memory-constrained QMARL can be effectively applied to perform complex operations such as high-resolution image processing and large-scale stereoscopic data processing.}

\section{Concluding Remarks}\label{sec:5}
In modern network research, autonomous mobility systems are widely used for pervasive large-scale management due to the development of MARL algorithms. However, MARL can be inefficient in terms of parameter utilization and convergence when large number of agents are utilized. To tackle this problem, based on quantum supremacy, QC-based MARL, i.e., QMARL, is designed based on the concept of actor-critic network under NISQ limitations for efficient parameter utilization, fast convergence, and scalability. Lastly, our QMARL can be further improved with PVM in terms of scalability, i.e., action dimension reduction into a logarithmic scale.



\bibliographystyle{IEEEtran}
\bibliography{ref_aimlab, ref_quantum}

\vspace{10mm}
{\small\textbf{Dr. Soohyun Park} is a postdoctoral scholar at Korea University, Korea. She received IEEE Vehicular Technology Society (VTS) Seoul Chapter Award (2019).}

\vspace{5mm}
{\small\textbf{Jae Pyoung Kim} is an M.S. student at Korea University, Korea.}

\vspace{5mm}
{\small\textbf{Chanyoung Park} is a Ph.D. student at Korea University, Korea.}

\vspace{5mm}
{\small\textbf{Prof. Soyi Jung} is an assistant professor at Ajou University, Korea.}

\vspace{5mm}
{\small\textbf{Prof. Joongheon Kim} is an associate professor at Korea University, Korea. 
He serves as an editor for \textit{IEEE Transactions on Machine Learning in Communications and Networking} and \textit{IEEE Communications Standards Magazine}. He received {IEEE Systems Journal} Best Paper Award (2020), IEEE ComSoc MMTC Outstanding Young Researcher Award (2020), and ComSoc MMTC Best Journal Paper Award (2021).}

\end{document}